\documentclass[referee]{aa}
\usepackage{epsfig}

\def\mearth{M_\oplus}

\def\msun{M_\odot}

\def\simgr{\,\hbox{\hbox{$ > $}\kern -0.8em \lower 1.0ex\hbox{$\sim$}}\,}
\def\simle{\,\hbox{\hbox{$ < $}\kern -0.8em \lower 1.0ex\hbox{$\sim$}}\,}
\def\beq{\begin{equation}}
\def\eeq{\end{equation}}

\def\simgr{\,\hbox{\hbox{$ > $}\kern -0.8em \lower 1.0ex\hbox{$\sim$}}\,}
\def\simle{\,\hbox{\hbox{$ < $}\kern -0.8em \lower 1.0ex\hbox{$\sim$}}\,}
\def\beq{\begin{equation}}
\def\eeq{\end{equation}}

\def\apj{ApJ}                 
\def\apjl{ApJ}                
\def\apjs{ApJS}               
\def\ao{Appl.Optics}          
\def\aap{A\&A}                
\def\aa{A\&A}                
\def\mnras{MNRAS}             

\def\({\left(}
\def\){\right)}
\def\<{\left<}
\def\>{\right>}

\begin{document}

\title{Formation and structure of the three Neptune-mass planets system around HD69830}

\author{Yann Alibert,$^{1}$ Isabelle Baraffe,$^{2}$ Willy Benz,$^{1}$ Gilles Chabrier,$^{2}$ \\
Christophe Mordasini,$^{1}$ Christophe Lovis,$^{3}$ Michel Mayor,$^{3}$ Francesco Pepe,$^{3}$ \\
Fran\c{c}ois Bouchy,$^{4}$ Didier Queloz,$^{3}$ \& Stephane Udry$^{3}$}

\institute{
$^{1}$ Physikalisches Insitut, University of Bern, Sidlerstrasse 5, CH-3012 Bern, Switzerland \\
$^{2}$ C.R.A.L., Ecole Normale Super\'ieure 46 all\'ee d'Italie, 69007 Lyon, France \\
$^{3}$ Observatoire de Gen\`eve 51 ch. Des Maillettes, 1290 Sauverny, Switzerland \\
$^{4}$ Institut d'Astrophysique de Paris 98bis Bd Arago, 75014 Paris, France}

\offprints{Yann ALIBERT, \email{yann.alibert@space.unibe.ch}}

\date{Received / Accepted}

\abstract{

Since the discovery of the first giant planet outside the solar system in 1995 (Mayor \& Queloz 1995), more than
180 extrasolar planets have been discovered. With improving detection capabilities, a new
class of planets with masses 5-20 times larger than the Earth, at close distance from their
parent star is rapidly emerging. Recently, the first system of three Neptune-mass planets
has been discovered around the solar type star HD69830 (Lovis et al. 2006). Here, we present and discuss a
possible formation scenario for this planetary system based on a consistent coupling between
the extended core accretion model and evolutionary models (Alibert et al. 2005a, Baraffe et al. 2004,2006). 
We show that the innermost planet formed from an embryo having started inside the iceline is composed
essentially of a rocky core surrounded by a tiny gaseous envelope. The two outermost planets
started their formation beyond the iceline and, as a consequence,  accrete a substantial
amount of water ice during their formation. We calculate the present day thermodynamical
conditions inside these two latter planets and show that they are made of a rocky core
surrounded by a shell of fluid water and a gaseous envelope.

\keywords{stars: planetary systems -- stars: planetary systems: formation}
}

\titlerunning{Formation and structure of the system around HD69830}
\authorrunning{Alibert Y. et al.} 
\maketitle

\section{Introduction}

The three Neptune-mass planetary system orbiting HD69830, a 4-10 Gyr old nearby 
star with a mass estimated at $0.86 \pm 0.03 \msun$, has been discovered 
through high precision measurements obtained with the HARPS spectrograph installed at La Silla, Chile (Lovis et al. 2006).
The three planets, planets b,c and d,
are located at 0.0785, 0.186 and 0.63 AU from the 
central star, and their minimum masses are equal to 10.2, 11.8, 18.1 $\mearth$  
respectively. 
This system, with three sub-Neptune mass planets within 1 AU, represents 
a considerable challenge for planet formation models, namely the disk instability (DI) model and the
core-accretion (CA) model.

\section{Formation by disk instability}

In the DI model, gravitational instabilities directly lead to
the formation of clumps that eventually evolve to form giant planets. In the case of
the present planetary system, this formation mechanism can be ruled out for two reasons.
First, the inner regions of the disk are too hot for gravitational instabilities to
take place. Second, gravitational instabilities at larger distances produce clumps
with masses much larger than those considered here (e.g. Boss 2001). Hence, even if subsequent migration
brings these clumps within 1 AU, they would be much more massive than the planets
considered here. Indeed, at least for the two outermost planets, we show that mass loss from
evaporation induced by the host star's high energy radiation is negligible.

It has been suggested that low mass planets could form in the framework of the DI model,
assuming the presence of a FUV/EUV source close to the formation site (e.g. a
\textit{close-in} O star) that would evaporate the gas envelope of initially
larger mass planets (Boss 2006). If such an external source is present,
photoevaporation indeed occurs at distances from the central star larger than $r_{\rm e} \propto G M_{\rm star} / c_s^2$,
where $M_{\rm star}$ is the mass of the central star, and $c_s$ is the sound speed
of gas heated up by the FUV/EUV flux (Boss 2006). Estimations of $r_{\rm e}$ depend
on the energy of the incoming flux, and range from 5 AU for EUV to 50 AU for FUV (Johnstone et al. 1998).
In the case of the Solar System, these values are consistent with a possible
evaporation of Uranus and Neptune, and the preservation of Jupiter's and Saturn's gaseous
envelopes. Given that the mass of HD69830 is close to solar, the photoevaporation radius
$r_{\rm e}$ can be expected to be similar as well. If this is correct, the photo-evaporation
of planet d located well inside $r_{\rm e}$ appears quite impossible. However,
other calculations point out that  $r_{\rm e}$ could be as low as 1 AU for EUV flux and 10 AU
for FUV flux (Adams et al. 2004). If these calculations are correct,  the evaporation of  planet d
by an external source would be marginally possible. However, in this case the survival of
Jupiter's atmosphere becomes a problem unless one argues that evaporation took place due to
\textit{EUV flux} in the case of the HD69830 system, and due to \textit{FUV flux} in the case
of the Solar System. Furthermore, we note that reconciling the DI/photoevaporation model and
the present system  requires evaporation to start \textit{after} the outermost planet has reached
a location close to its present location (inward migration of planets formed by DI at large
distances requires a substantial amount of gas). Finally, at a given EUV flux, a planet evaporates
increasingly faster as its mass diminishes (Baraffe et al. 2004,2006). Hence, to account for  planet d requires
a very fine tuning between the start and stop of the evaporating flux. While not impossible,
such special circumstances appear quite unlikely.  Finally, we note that DI has been also strongly
excluded as a possible formation mechanism of the HD149026 system (Sato et al. 2005).
The discovery of a Jupiter mass planet orbiting at yet larger distances from HD69830 would 
definitely rule out the DI/photoevaporation as a possible formation mechanism, at least for this 
particular system. 

\section{Formation by nucleated instability}

In the CA model, a solid core is first formed by the accretion of solid
planetesimals. When its mass is large enough, it can accrete gas in a runaway process, 
rapidly building up a giant planet (Alibert et al. 2005a, Hubickyj et al. 2005).
In the framework of this model, the \textit{in-situ} formation of cores large enough to trigger
the runaway gas accretion, and consequently the \textit{in-situ} formation of \textit{close-in} giant
planets, is prevented by the sheer lack of solid material that close to the star. On the one hand,
low mass disks simply lack the necessary amount of solids, while on the other hand more
massive disks with similar lifetimes are too hot for solids to condense at these short
distances. Thus, to reach their present mass, the planets orbiting HD69830 must have
swept planetesimals over distant regions of the disk. Therefore, the discovery of this
system of hot-Neptune planets implies (if there were any doubts left) that significant
planetary migration had to occur.

In order to compute the formation and the evolution of this system,
we use the extended CA model which takes into account the migration of the proto-planets as well as the
evolution of the disk. We also consider the evolution of the new born planets to the present day by taking
into account the  effects of irradiation and evaporation due to stellar
radiation. Our entire approach has been extensively described elsewhere (Alibert et al. 2005a, Baraffe et al., 2004,2006),
where the reader is referred for more details. 
Using these models, we performed a large
number (few tens of thousand) of simulations to find all initial conditions leading to a planetary system comparable to this one.
Assuming a central star of 0.86 $\msun$ and  a dust-to-gas ratio of $1/70$, accounding for the slightly sub-solar metallicity 
of the star (Lovis et al. 2006), we start
our calculations with a protoplanetary disk and seed the three planets by means of
three embryos of 0.6 $\mearth$ each.
We explore the different disk characteristics (mass and lifetime) and initial locations of
the three embryos leading to the observed characteristics of the three planets.
Finally, note that we do not take into account gravitational interactions between the three forming
planets that could alter the migration rates. However, we have checked that the planets do not cross 
their mutual main mean motion resonances during the formation (in particular the 1:2, 1:3
and 2:3 resonances).

Since planets migrate significantly during their growth, they eventually encounter
the wake created by the preceeding planet. To account for this, we considered the
formation of the planets from inside out, each embryo being started at a different
distance and time and followed while migrating through a disk already modified by
the preceding planets. As the planet encounters a region depleted of planetesimals
by the passage of a previous body, the accretion rate of solids is vanishing, leading to the
suppression of the main heating source and the planet can accrete gas (Alibert et al. 2005b)
 at a rate essentially given by its Kelvin-Helmholtz timescale (Ida \& Lin 2004).
Note that the absence of solids in the innermost regions of the disk
(inside $\sim$0.35 AU), where the temperature exceeds the evaporation temperature of
silicates ( $\sim$1600 K), leads to a similar effect. 

\begin{figure}
\begin{center}
\epsfig{file=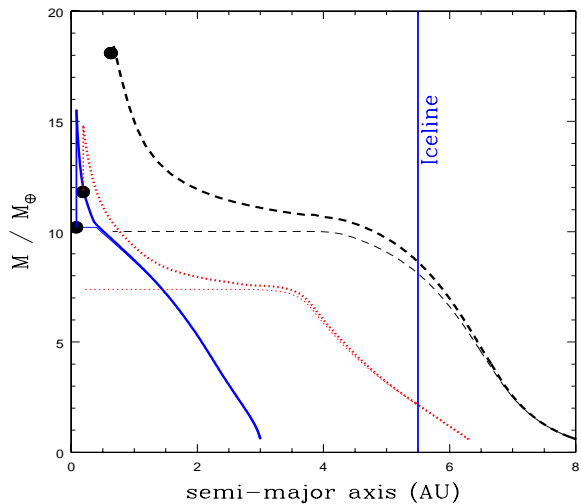,height=70mm,width=80mm}
\end{center}
\caption{Formation/evolutionary tracks of the planetary system orbiting HD69830.
The total mass (thick lines) and the core mass (thin lines) for the three planets
are given as a function of semi-major axis which is decreasing with time as a
result of migration. The iceline is indicated by the vertical line. The minimum
mass and semi-major axis derived from the observations are  indicated as big dots.
The solid lines correspond to the innermost planet, the dotted lines to the middle one,
and the dashed lines to the outermost one. The vertical lines at 0.08 AU and 0.18 AU
reflect the evaporation of the two innermost planets during 4-10 Gyr, the estimated age of
the system.
}
\label{fig1}
\end{figure}

The formation and evolutionary tracks of the three planets in a mass versus semi-major
axis diagram are shown in Figure \ref{fig1}. We found that, in order to reproduce the mass
and the semi-major axis of these three planets, gas disks surface densities around
800 g/cm$^2$  at 5 AU are required. This corresponds to a disk mass (between 0.07 AU
and 30 AU,  and assuming an initial power law for the disk surface density
$\Sigma \propto r^{-3/2}$) of 0.07 $\msun$ and disk lifetime of about 2 Myr, both values compatible
with values inferred from observations (Haisch et al. 2001, Beckwith \& Sargent 1996). 
The innermost planet starts well inside the iceline and grows by accreting essentially rocky 
planetesimals and gas. On the contrary, the two outermost planets start beyond the iceline, 
and  accrete a significant amount of icy planetesimals (as well as gas and rocky planetesimals).

Planet b's embryo starts at 3 AU, and, at the time it enters the innermost regions of the disk (below 0.35 AU), 
the accretion rate of solids drops dramatically triggering the accretion of gas. The planet 
reaches its final position at the time the disk vanishes and consists, at the end of the 
formation process, of a solid core of $\sim$10 $\mearth$ surrounded by an envelope of 
$\sim$5 $\mearth$. Planet c's embryo starts at 6.5 AU and accretes planetesimals until
it enters the region already depleted by the innermost planet (3AU). Again, this depletion
triggers the accretion of gas, leading to a planet consisting in a rocky/icy core of  
$\sim$ 7.5 $\mearth$ ($\sim$5.7 $\mearth$ of rocks and $\sim$1.7 $\mearth$ of ices, assuming a standard
ices-to-rocks ratio of 4) and a H/He gaseous envelope of $\sim$ 7.5 $\mearth$.
Finally, planet d's embryo starts at $\sim$ 8 AU (well beyond the iceline) and 
accretes a large amount ($\sim$ 60 \%) of icy planetesimals.  At the time the growing planet 
enters the region of the disk already depleted by the second planet, gas accretion is again 
triggered. The final planet consists, at the end of the formation process, of a $\sim 10 \mearth$ 
core ($\sim 5.2 \mearth$ of rocks and $\sim 4.8 \mearth$ of ices), surrounded by a gaseous envelope 
of $\sim 8 \mearth$.

Starting from these three formation models, we have then followed the evolution
of the three planets taking into account evaporation and irradiation effect (Baraffe et al. 2004,2006)
during 4 to 10 Gyr, which corresponds to the estimated age of the HD69830 system.

\section{Evolution and evaporation}

For this calculation, we took the initial internal compositions obtained at the end of the
formation phase. The cores of the two innermost planets are assumed to be made of
dunite (Mg$_2$SiO$_4$ - the amount of ice in the core of the middle planet is negligible
for the evolution calculation). For the outermost one, we have considered the two limiting cases,
one with a  pure icy core, and one with a pure rocky core. Each of the three planets
is surrounded by a hydrogen/helium envelope whose mass is provided by
the formation model.
We have taken into account the effect of the incident radiation of the parent star,
which modifies the internal structure and the cooling rate of close-in planets, as
well as the mass loss due to the evaporation of the outermost layers of the planet's envelope heated
by the incident stellar high energy flux. The evaporation rate was chosen to be 1/20
the maximal escape rate of Lammer et al. (2003), a value obtained by various recent detailed
hydrodynamical calculations (Tian et al. 2005, Yelle 2004), and consistent with lower limits inferred from
observations (Vidal-Madjar et al. 2003).

The effect of irradiation and evaporation is found to be completely negligible for
the planet d and to lead to only a small (5\%-10\%) mass loss for planet c (see vertical line
at 0.18 AU in Fig. \ref{fig1}). For planet b, however, it is significant (vertical line at 0.08 AU in
Fig. \ref{fig1}). Within a few Gyr, essentially all its envelope is evaporated, leaving behind a solid core with
only a tiny (less than 2 $\mearth$) gaseous atmosphere. Calculations done with slightly
different initial conditions (core and envelope mass), yield similar results. The
radius of  a rocky core (dunite) of 10 $\mearth$ is $0.18 R_{\rm J}$ (radius of Jupiter).
which gives a lower limit for the expected radius of this planet,
In the case it was able to retain even a tiny atmosphere its radius will be larger:  for an envelope
mass of $\sim 2 \mearth$, the radius would be increased to $\sim 0.45 R_{\rm J}$.
The radii of the two outermost planets are found to be 50-60\% $R_{\rm J}$, depending
on the precise composition of the planet's core and the age of the system.

\begin{figure}
\begin{center}
\epsfig{file=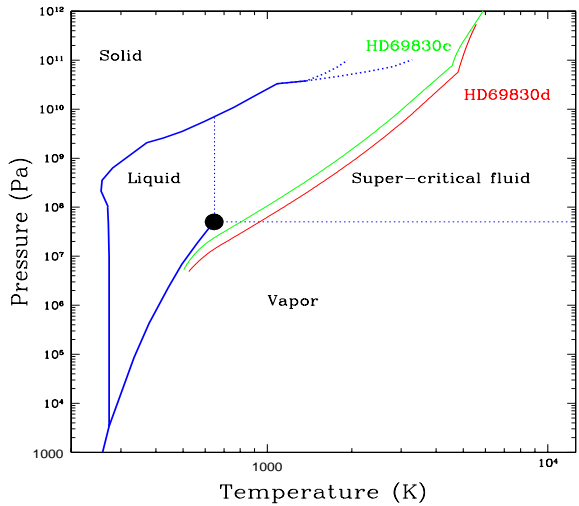,height=70mm,width=80mm}
\end{center}
\caption{Thermodynamical conditions inside planets c and d
and simplified phase diagram of water. The big dot indicates the
position of the critical point. The two heavy dotted curves at high pressures give
the likely location of the melting curve (Lin et al. 2005).
The thermodynamical conditions are calculated after 4 Gyr, assuming a dunite
core for HD69830c, and a ice core for HD69830d. The kink around T $\sim$ 4000-5000 K in the two
internal profiles indicates the envelope/core transition (the core is characterised by higher pressures).
The two planets harbour a similar structure: a central rocky core, surrounded by a shell
of super-critical fluid water, and a hydrogen/helium gas envelope. The results after 10 Gyr,
or assuming a dunite core for HD69830d are similar and are not presented for
clarity. 
}
\label{fig2}
\end{figure}

Figure \ref{fig2} shows the thermodynamical conditions inside planets c and d,
after 4 Gyr, the minimum age of the system, (the results after 10 Gyr 
are very similar), together with a simplified phase diagram of water.
For these two planets,  the temperature and pressure are such
that water is likely to exist under the form of a super-critical fluid.
However, note that no experimental data regarding melting of water are
available at the high temperatures characteristic of the planet's interiors.

\section{Conclusions}

We have presented calculations which provide a fully consistent scenario for 
the formation and evolution of the planetary system around HD69830.
From the calculations presented we can infer the following general scenario for the
formation of the system. All three planets start by accreting
planetesimals and very little gas as they migrate inwards until they reach a
region depleted in solids either by the passage of a previous planet or because
of too high temperatures. The main heating source being suppressed they essentially
accrete gas at a rate given by their Kelvin-Helmholtz (KH) timescale (Ida \& Lin 2004). To remain
of Neptune-mass without requiring unlikely timing with the disapearence of the disk, a given planet must enter
this depleted region when its KH timescale is of the order of the lifetime of the disk which
corresponds to a mass of order ~8-12 $\mearth$ (Ida \& Lin 2004). For the three planets to collect
this mass of heavy elements implies a significant amount of migration of the growing
cores. 

A question which naturally arises with such a planet formation model, is the degree of fine-tuning 
of the initial conditions needed to produce a planetary system with similar properties 
(an exact match is meaningless). In this regard, the protoplanetary disk mass and 
lifetime we require are typical of observed values (Haisch et al. 2001, 
Beckwith \& Sargent 1996). In fact, the major constraint comes from the fact 
that the planets, in order to remain of small mass, must enter the planetesimal 
depleted region of the disk at a time when their accretion timescale 
(roughly the core's Kelvin-Helmholtz timescale) is comparable to the 
lifetime of the disk. For the three planets (b,c and d), this timescale 
is around  1 Myr, 0.6 Myr and 0.2 Myr. In our simulation the lifetime 
of the disk is of order 2 Myr. Hence, it is only for the second and 
third planet (c and d) that this requirement is really limiting the 
possibilities, but it is certainly not fine-tuning.

For the first time, our consistent formation/evolution calculations lead to the
determination of the bulk composition and the inner structure of three Neptune-mass planets:
the innermost one consists of a rocky core, with possibly a tiny gaseous envelope,
whereas the two outermost planets are made of a central rocky core, a shell of
super-critical fluid water and a gaseous envelope.
A clear test of the present formation and evolution scenario could be achieved
by the determination of the mean density of the planets. This would only be possible if the
system is seen edge-on and transits are detected so as to measure the radius of the planets. While
difficult from the ground, such observations are within reach of HST, COROT or KEPLER.
Even if the present system does not lead to observable transits, it is likely that similar,
transiting Neptune-mass systems will be discovered in a near future. Confrontation of the
present theory with such observations will improve dramatically our understanding of planet formation.

Finally, Spitzer observations of the HD69830 system have revealed the presence of micron sized
dust at distances lower than 1 AU from the central star, that could result from
the presence of an asteroid belt (Beichman et al. 2005).
Preliminary order of magnitude estimates have shown that the passage of the two inner planets
during their formation may significantly but not completely deplete the asteroid belt. Hence,
the belt,  if present prior to the formation of the planets, would be able to survive at least
in part. Interestingly, we note that dust is observed in regions in mean motion resonances
with the outermost planet (1:2 and 1:3), that may excite the asteroids, leading to collisions
and dust production.

\acknowledgements

This work was supported in part by the Swiss National Science Foundation.

\end{document}